\documentclass[prd,12pt]{revtex4}

\usepackage{amssymb,amsthm}
\usepackage{epsfig}

\begin{document}

\title{Halos of Modified Gravity\footnote{This essay was selected for Honorable
Mention in the annual essay competition of the Gravity Research Foundation for the year
2008.}}

\author{Kirill Krasnov\,$^a$}\email{kirill.krasnov@nottingham.ac.uk}
\author{Yuri Shtanov\,$^b$}\email{shtanov@bitp.kiev.ua}
\affiliation{$^a$School of Mathematical Sciences, University of
Nottingham, Nottingham, NG7 2RD, UK. \\
$^b$Bogolyubov Institute for Theoretical Physics, Kiev 03680, Ukraine.}

\date{March 28, 2008}

\begin{abstract}
We describe how a certain simple modification of general relativity, in which
the local cosmological constant is allowed to depend on the space-time
curvature, predicts the existence of halos of modified gravity surrounding
spherically-symmetric objects.  We show that the gravitational mass of an
object weighed together with its halo can be much larger than its gravitational
mass as seen from inside the halo.  This effect could provide an alternative
explanation of the dark-matter phenomenon in galaxies.  In this case, the local
cosmological constant in the solar system must be some six orders of magnitude
larger than its cosmic value obtained in the supernovae type Ia experiments.
This is well within the current experimental bounds, but may be directly
observable in the future high-precision experiments.
\end{abstract}

\maketitle

The long-standing problem of gravitationally detectable but otherwise
unobserved dark matter can, in fact, suggest that general relativity theory
needs to be modified under certain physical conditions which are not
encountered in the solar system.  This case has become particularly strong
after the discovery --- made by Milgrom in 1983 and confirmed afterwards ---
that the need for dark matter in galaxies arises as soon as the Newtonian
acceleration of test bodies reaches a tiny universal value $a_0 \simeq 2 \times
10^{-10}$~m/s$^2$, which happens to be of the order of $c H_0$, where $H_0$ is
the current Hubble parameter, and $c$ is the speed of light \cite{Milgrom}.
Universality of this kind can hardly be explained in the dark-matter paradigm,
in which the relative amount and distribution of dark and visible matter can
vary significantly from object to object, reflecting the haphazard history of
formation of individual gravitationally bound systems \cite{Milgrom1}.

In general relativity, Newtonian acceleration itself is not a locally
observable quantity, in view of the strong equivalence principle inherent in
this theory.  Therefore, in order to substantiate the principles of modified
gravity and to construct a self-consistent relativistic theory exhibiting
effects of dark matter, one has either to introduce additional special vector
fields, with respect to which acceleration can be defined, or to choose another
invariant indicating the conditions under which gravity is to be modified.  The
first approach is realized in the tensor--vector--scalar theories of modified
gravity \cite{Bekenstein:2004ne}; the second approach is implemented, in
particular, in the modified theory of gravity which is the subject of this
essay.

A general-relativistic quantity most closely connected with Newtonian relative
acceleration is, of course, tidal acceleration.  For two test bodies at a small
relative proper distance $d$ in an external gravitational field, the relative
acceleration is proportional to $d$:
\begin{equation}
\ddot d \equiv a_{\rm tidal} = \beta d \, .
\end{equation}
This is what is called tidal acceleration, and the constant of proportionality
$\beta$ is part of the curvature tensor, characterizing the external
gravitational field.  It looks reasonable to ask whether one can construct a
mathematically consistent theory in which curvature invariants play the role of
indicators signalling the breakdown of general relativity. At first sight, such
a modification would require terms in the action involving higher powers of the
curvature tensor, resulting in a higher-order metric theory of gravity with its
extra degrees of freedom and severe problems of instability. It is surprising,
therefore, that a formulation of general relativity exists in which the
suggested modifications can easily be implemented without introducing any new
degrees of freedom (in particular, without increasing the order of differential
equations).

The theory which is the subject of this essay is a slight modification of the
less known, albeit rather old, formulation of general relativity in the
language of self-dual two-forms due to Pleba\'nski \cite{Plebanski}. The
Pleba\'nski action can be written as
\begin{equation} \label{es-Plebanski}
S = {1 \over 8 \pi G} \int \mbox{Tr} \left[ B F + B \left( \Psi - \frac13
\Lambda \right) B \right] \, ,
\end{equation}
where the $\mathfrak{su}(2)$-Lie-algebra valued two-form $B$ carries
information about the space-time metric, and $F$ is the curvature of an
$\mathfrak{su}(2)$ connection $A$. The quantity $\Psi$ is a symmetric traceless
$3 \times 3$ matrix, which is the Weyl curvature in this formulation.  Wedge
products of forms is assumed in (\ref{es-Plebanski}), and the symbol $\Lambda$
stands for the usual cosmological constant.  A remarkable feature of
formulation (\ref{es-Plebanski}) of general relativity is that the Weyl
curvature $\Psi$ enters here as a Lagrange multiplier.  This enables one to
consider a simple class of modifications of action (\ref{es-Plebanski}) in
which the cosmological constant is allowed to become a function of the two
$\mathfrak{su}(2)$ invariants of $\Psi$, which are $\mbox{Tr}\, \Psi^2$ and
$\mbox{Tr}\, \Psi^3$:
\begin{equation} \label{es-lambda}
\Lambda = \Lambda \left( \mbox{Tr}\, \Psi^2, \mbox{Tr}\, \Psi^3 \right) \quad
\mbox{in action (\ref{es-Plebanski})} \, .
\end{equation}
This change, in fact, naturally arises as a modification of the classical
action (\ref{es-Plebanski}) by quantum corrections \cite{paper1}.
\begin{figure}
\begin{center}
\epsfig{figure=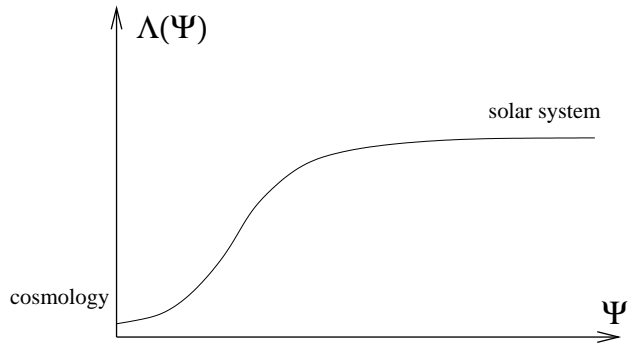,width=0.5\textwidth}
\end{center}
\caption{Regions of relatively high curvature (solar system) correspond to
large local values of $\Lambda$. Regions of small (zero) curvature have local
values of $\Lambda$ of relevance for cosmology. \label{fig:1}}
\end{figure}

On the classical level, the ``cosmological function'' $\Lambda \left(
\mbox{Tr}\, \Psi^2, \mbox{Tr}\, \Psi^3 \right)$ can be chosen arbitrarily: any
choice gives a consistent theory propagating just two degrees of freedom, as is
the case in general relativity \cite{Krasnov:2007cq}.  Therefore, one can use
the opening freedom to specify it in a desirable way.  First of all, in accord
with the preceding reasoning, we would like this function to vary
insignificantly in space-time regions of sufficiently high values of the Weyl
curvature, such as those encountered in solar neighborhood, in order to comply
with the solar-system experiments which show no significant deviation from the
general-relativistic predictions. Next, we can also reasonably assume that this
function is smooth in the neighborhood of zero curvatures. The limit of this
function as $\Psi \to 0$ is to be associated with the value of the cosmological
constant observed in the supernovae type Ia experiment.  Indeed, in the ideal
cosmological background, the quantity $\Psi$ vanishes due to space-time
symmetry; in the real universe, its value will be close to zero and will
determine the effective value of cosmological constant via (\ref{es-lambda}).
The function $\Lambda \left( \mbox{Tr}\, \Psi^2, \mbox{Tr}\, \Psi^3 \right)$
will then smoothly interpolate between these two regions of small and large
values of $\Psi$; see Fig.~\ref{fig:1}.

Many of the physical implications of our theory can be deduced from the
spherically symmetric vacuum solution, which was obtained in our paper
\cite{paper2}. Notably, the solution respects the analog of the Birkhoff
theorem saying that it is static. Due to spherical symmetry, the traceless
symmetric $3 \times 3$ matrix $\Psi$ is described by a single function $\psi
(r) $:
\begin{equation}
\Psi^{ij} = \psi (r) \left( 3 {x^i x^j \over r^2}  - \delta^{ij} \right) \, ,
\end{equation}
where $x^i$ are the Euclidean spatial coordinates, and $r^2 = \sum\limits_i
\left(x^i\right)^2$.  In this case, the cosmological function (\ref{es-lambda})
of two invariants becomes a function of $\psi$ only, and its derivative with
respect to $\psi$, which we denote by $\Lambda_\psi$, represents a
dimensionless quantity characterizing the deviation of the theory from the
general-relativistic behavior: the condition $\left| \Lambda_\psi \right| \ll
1$ implies the validity of general relativity.

One of the consequences of the specific form of the cosmological function
$\Lambda$ described above is that an isolated spherically symmetric body in our
theory is surrounded by a region of approximate validity of general relativity,
in which the Weyl curvature is large, and the condition $\left| \Lambda_\psi
\right| \ll 1$ is well satisfied. At large distances from such a body, the
value of $\psi$ tends to zero, and the derivative $\Lambda_\psi$ becomes small
again since this function smoothly depends on the two invariants which are,
respectively, quadratic and cubic in $\psi$. Thus, the solution is
asymptotically general-relativistic at large radial distances as well as at
small ones. However, at intermediate distances at which $\Lambda_\psi$ is not
much smaller than unity, the theory can significantly deviate from general
relativity. It is this region around a central body that we call {\em halo of
modified gravity\/}.
\begin{figure}
\begin{center}
\epsfig{figure=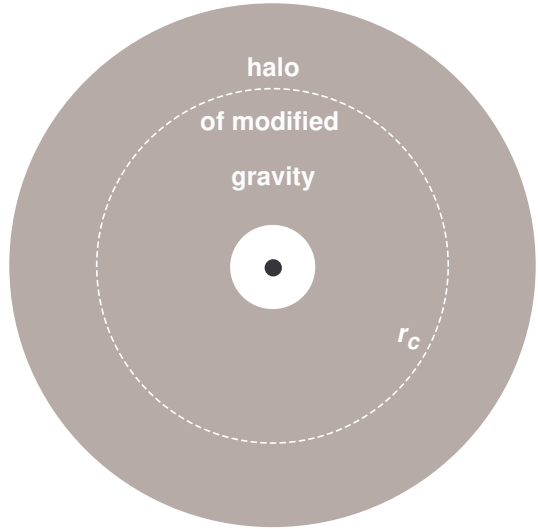,width=0.45\textwidth}
\end{center}
\caption{Halo of modified gravity for the $\Lambda$-function
(\ref{es-example}), drawn to scale according to (\ref{es-numeric}).
\label{fig:2}}
\end{figure}

The two-form field $B$ in the theory under consideration bears information
about the space-time metric. It can be deduced from the requirement that the
two-form $B$ is self-dual in this metric and that its volume form coincides
with $\frac13 {\rm Tr}\, \left( B \wedge B \right)$. This leads to the
expression for the metric in the form
\begin{equation} \label{es-metric}
ds^2 = { 1 \over \sqrt{1 - \Lambda_\psi / 3 }} \left[ f^2 (r) dt^2 - g^2 (r)
dr^2 - r^2 d \Omega^2 \right] \, ,
\end{equation}
where the functions $f(r)$ and $g(r)$ are specified by
\begin{equation} \label{es-relations}
g^2 = \left({1 - \Lambda_\psi / 3 \over 1 + \Lambda_\psi / 6}\right) g_*^2 \, ,
\qquad f^2 = \left({1 - \Lambda_\psi / 3 \over 1 + \Lambda_\psi / 6}\right) Z^2
(r) g_*^{-2}
\end{equation}
and
\begin{equation} \label{es-grs}
g_*^{-2} = 1 - {r_s (r) \over r} - \frac13 \Lambda (r) r^2 \, , \qquad r_s (r)
= { r_s^{(\infty)} \over Z(r) } \, .
\end{equation}
The relation between $\psi (r)$ and effective Schwarzschild radius $r_s (r)$ is
usual:
\begin{equation} \label{es-psi}
\psi (r) = {r_s (r) \over 2 r^3} \, ,
\end{equation}
and the function $Z(r)$ is implicitly expressed through $\psi (r)$ as follows:
\begin{equation}
Z(\psi) = \exp \left( \int\limits_0^\psi {\Lambda_\psi (\psi') \over 6 \psi'} d
\psi' \right) \, .
\end{equation}
The constant $r_s^{(\infty)}$ then corresponds to the Schwarzschild radius at
spatial infinity.

To be more specific, consider a concrete example of the function $\Lambda
(\psi)$:
\begin{equation} \label{es-example}
\Lambda (\psi) = \Lambda_0 + {3 \alpha \over \ell^2} \log \left[ 1 + \left(
\ell^2 \psi \right)^2 \right] \, ,
\end{equation}
where $\ell$ is a constant of dimension length, and $\alpha$ is a dimensionless
parameter.  If $\alpha \ll 1$, then $\left| \Lambda_\psi \right| \ll 1$
everywhere, and the halo of modified gravity is absent.  However, there is also
a theoretical upper bound $\Lambda_\psi < 3$ which, if violated, results in
unwanted singularities in the theory, as is clear from (\ref{es-metric}),
(\ref{es-relations}). As the maximum value of $\Lambda_\psi$ for function
(\ref{es-example}) is equal to $3 \alpha$, it is interesting to consider the
values $\alpha \lesssim 1$. Then the halo of modified gravity is characterized
by the magnitudes of $\ell^2 \psi$ in the range $(6 \alpha)^{-1} \lesssim
\ell^2 \psi \lesssim 6 \alpha$.  This relation determines the radial extension
of the halo for every object, which, in view of (\ref{es-psi}), is given by
\begin{equation} \label{es-halo}
\left(12 \alpha Z_0 \right)^{- 1/3} \lesssim {r \over r_c} \lesssim \left( 3
\alpha \right)^{1/3} \, , \qquad r_c = \left(r_s^{(\infty)} \ell^2
\right)^{1/3} \, ,
\end{equation}
where
\begin{equation} \label{es-z0}
Z_0 = \lim_{\psi \to \infty} Z (\psi)  = e^{\alpha \pi / 2} \, .
\end{equation}
Taking the maximal values of $\alpha \simeq 1$, we get an estimate
\begin{equation} \label{es-numeric}
0.26 \lesssim {r \over r_c} \lesssim 1.44 \, ,
\end{equation}
which describes the relative extension of the halo of modified gravity in our
specific example (see Fig.~\ref{fig:2} for its depiction).

Looking at (\ref{es-relations}) and (\ref{es-grs}), one can notice two
interesting features characterizing the spherically symmetric solution: (i)~the
effective Schwarzschild radius $r_s$ becomes distance dependent, being
inversely proportional to $Z (r)$; (ii)~the $g_{00}$ component of the metric
acquires an additional redshift/blueshift factor $Z^2 (r)$. Note that, for any
reasonable choice of the $\Lambda$-function, the function $Z(r)$ is
monotonically decreasing with distance, asymptotically approaching unity at
large distances.

The first effect can be a modified-gravity substitute for dark matter in
galaxies. Indeed, it means that any object weighed from infinity, i.e.,
together with its halo, is much more massive than how it looks from the inside
of its halo. However, this is precisely the phenomenon motivating introduction
of dark matter in galaxies and galaxy clusters. The theory under consideration
exhibits this dark-matter phenomenon as a consequence of modification of
gravity.

To give an estimate of parameters that would be necessary here, we take $\alpha
\lesssim 1$, obtaining the effect of gravitational mass increase by a factor
$Z_0 = e^{\pi / 2} \approx 4.8$ --- typical for the dark-to-luminous matter
ratios in galaxies. If needed, a larger mass ``magnification'' is possible by
choosing a different form of the $\Lambda$-function. A typical representative
of a situation requiring dark matter is a spiral galaxy like our Milky Way, of
mass $M_g \sim 10^{11} M_\odot$, in which deviations from Newton's behavior
(flat rotation curves) begin at distance $r_g \simeq 3$~kpc from the center.
Then relation (\ref{es-halo}) gives us the estimate
\begin{equation} \label{es-ell}
\ell \simeq \sqrt{12\, r_g^3 \over r_s} \simeq 5.7~\mbox{Mpc} \, ,
\end{equation}
where $r_s = 2 GM_g \simeq 10^{-2}$\,pc is the Schwarzschild radius associated
with the galaxy mass contained within the radius $r_g$.  Of course, this
estimate is rather crude. The real situation will be more complicated because
the halo of modified gravity for a galaxy, formed by many stars, will not have
a spherical shape.  However, it demonstrates that, in the theory under
investigation, the usual halo of dark matter might in principle be replaced by
the halo of modified gravity, to the same effect.

The described mechanism of mass ``magnification'' implies that the values of
the local ``cosmological constant'' should be different deep inside and outside
of the halo of modified gravity. Turning again to our example
(\ref{es-example}), we see that its value near the sources of gravity can be
many orders of magnitude larger than the cosmological value $\Lambda_0$. This
is true for our estimate (\ref{es-ell}) of the parameter $\ell$, which gives
\begin{equation} \label{es-lambdasol}
\Lambda \simeq \Lambda_0 + {3 \alpha \over \ell^2} \simeq 10^6 \Lambda_0 \, .
\end{equation}
The current upper bounds on the cosmological constant in the solar system are
$\Lambda < 10^{10} \Lambda_0$, much higher than (\ref{es-lambdasol}), but the
future experiments can improve the precision and possibly be capable of
detecting the cosmological constant of this magnitude \cite{bounds}.

Another feature of the spherically symmetric vacuum solution that we mentioned
above is an unusual redshift/blueshift factor in the $g_{00}$ metric
coefficient, which is seen in equation (\ref{es-relations}).  The presence of
this factor, in particular, implies that the a photon emitted from a region of
high curvature will get blueshifted as it travels through the halo into a
region of low curvature.  An increase in the photon energy occurs exactly in
the same proportion as an increase of the gravitational mass of the central
body, since one and the same factor $Z(r)$ controls the redshift/blueshift in
(\ref{es-relations}) and the effective Schwarzschild mass in (\ref{es-grs}).
What makes this effect practically unobservable is the circumstance that
photons are usually emitted and detected in regions of high curvature, so that
their initial blueshifts are compensated by subsequent redshifts of exactly the
same magnitude.

The effects of increase of gravitational mass and photon's energy can be
unified in a simple physical picture.  As any entity propagates through space
from a region of high curvature (and high local cosmological constant) into a
region of low curvature (and low local cosmological constant), it is
essentially sliding along the ``potential well'' formed by the ``cosmological
function'' $\Lambda$.  This causes its gravitating energy to increase, and this
is what we observe: massive bodies develop the low-curvature halos of modified
gravity enhancing their weight, and photons get blueshifted in exactly the same
proportion.

The theory under consideration with a variable curvature-dependent cosmological
``constant'' exhibits many other interesting physical effects, which remained
beyond the scope of this short essay.  Its basic direct suggestion, which we
elaborated upon above, is that the value of the cosmological constant in the
solar system may be many orders of magnitude larger than the value measured
in cosmology. Whether this possibility is realized in Nature can only be
decided by experiment. However, the described theory of gravity already
represents an exciting playground for the idea that the dark-matter
phenomenon is an effect of modified gravity.

This work was supported by the European Science Foundation in frames of the
programme ``Quantum Geometry and Quantum Gravity.'' K.~K.\@ was also supported
by an EPSRC Advanced Fellowship, and Yu.~S.\@ was supported by the
``Cosmomicrophysics'' programme and by the Program of Fundamental Research of
the Physics and Astronomy Division of the National Academy of Sciences of
Ukraine.

\end{document}